\documentclass[%
aps,%
12pt,%
final,%
notitlepage,%
oneside,%
onecolumn,%
nobibnotes,%
nofootinbib,%
superscriptaddress,%
noshowpacs,%
showkeys,%
centertags%
]{revtex4}
\usepackage{amsfonts,amssymb,amsbsy}
\usepackage{graphicx,amsfonts}
\usepackage{amsfonts,amssymb,amsbsy}
\usepackage{array}
\usepackage{amsmath}
\usepackage{graphicx,amsfonts}
\usepackage{amsfonts,amssymb,amsbsy}

\begin{document}

\title{Discriminating an Arbitrary Number of Pure Quantum States by the Combined $\mathcal{CPT}$ and Hermitian Measurements
}

\author{Yaroslav Balytskyi}
\email{ybalytsk@uccs.edu}
\affiliation{Department of Physics and Energy Science, University of Colorado at Colorado Springs, Colorado Springs, CO 80933, USA
}
 
\author{Sang-Yoon Chang}
\email{schang2@uccs.edu}
\affiliation{Department of Computer Science, University of Colorado at Colorado Springs, Colorado Springs, CO 80933, USA
}
 
\author{Anatoliy Pinchuk}
\email{apinchuk@uccs.edu}
\affiliation{Department of Physics and Energy Science, University of Colorado at Colorado Springs, Colorado Springs, CO 80933, USA
}

\author{Manohar Raavi}
\email{mraavi@uccs.edu}
\affiliation{Department of Computer Science, University of Colorado at Colorado Springs, Colorado Springs, CO 80933, USA
}

\begin{abstract}
If the system is known to be in one of two non-orthogonal quantum states, $|\psi_1\rangle$ or $|\psi_2\rangle$, $\mathcal{PT}$-symmetric quantum mechanics can discriminate them, \textit{in principle},  by a single measurement.

We extend this approach by combining $\mathcal{PT}$-symmetric and Hermitian measurements and show that it's possible to distinguish an arbitrary number of pure quantum states by an appropriate choice of the parameters of $\mathcal{PT}$-symmetric Hamiltonian. 

\end{abstract}
\keywords{Quantum state discrimination;  $\mathcal{PT}$-symmetry; Unstructured database search;}

\maketitle

Quantum state discrimination is an important problem which has many applications in quantum computing and quantum cryptography; see a  review article, Ref.\cite{Rev}. In  quantum computation it is known that an unstructured database search can be mapped to the problem of distinguishing exponentially close quantum states, Ref.\cite{Exp}. In turn, it is related to the hash functions widely used in cryptography, Ref.\cite{Hash}.

Suppose, the system may be in one of two possible states, $|\psi_1\rangle$ or $|\psi_2\rangle$ and the task is to find in which of them the system is. A simple classical example: The experimenter is told that the coin is either fair or biased. The task is to determine by tossing the coin which option is true. 

Quantum state discrimination consists of finding an optimal observable and strategy of measurements, Ref.\cite{Rev}. In an conventional quantum mechanics, it's impossible to find the state of the system by a single measurement. However, it is not possible in general to repeat the measurement since it changes the state of the system.  Therefore, to know the state of the system with a high confidence level, one has to prepare a large number of identical samples. 

However,  $\mathcal{PT}$-symmetric quantum mechanics, Refs\cite{NH1, NH2, NH3}, offers new opportunities. If the $\mathcal{PT}$ symmetry of a Hamiltonian is not broken, then its eigenvalues are real and Hamiltonian determines an inner product giving an extra degree of freedom in comparison with a conventional quantum mechanics.

While it's an open question on how to properly treat the boundary between Hermitian and non-Hermitian regimes, Refs.\citep{Doubt1, Doubt2, Doubt3}, it's worth  considering a potential implications of an additional degree of freedom.

In particular, it was shown that in a $\mathcal{PT}$-symmetric quantum mechanics it's possible \textit{in principle} to discriminate between 2 states just by a \textit{single} measurement, Ref.\cite{PTDis}. On practice, discriminating  two non-orthogonal states in a closed system with one measurement can only be done with probability less then one, otherwise it would be a violation of unitarity. Therefore, even in the ideal noiseless case, sometimes we may have to apply the measurement more then once.

Two alternative solutions were proposed:
\begin{itemize}
\item \textit{Solution 1: Finding a $\mathcal{PT}$-symmetric Hamiltonian whose inner product interprets $|\psi_1\rangle$ and $|\psi_2\rangle$ as being orthogonal under the $\mathcal{CPT}$ scalar product.}
\item \textit{Solution 2: Finding a $\mathcal{PT}$-symmetric Hamiltonian under which the states $|\psi_1\rangle$ and $|\psi_2\rangle$ evolve into orthogonal states under Hermitian scalar product.}
\end{itemize}

We propose to combine \textit{Solution 1} and \textit{Solution 2} to extend this approach to be able to discriminate more than two states. 

Start with 3 states:

\begin{equation}
|\psi_1\rangle = 
\begin{pmatrix}
 \cos\left(\frac{\pi - 2\epsilon}{4}\right)\\
 -i \sin\left(\frac{\pi - 2\epsilon}{4}\right)
\end{pmatrix}; \\ |\psi_2\rangle = 
\begin{pmatrix}
 \cos\left(\frac{\pi + 2\epsilon}{4}\right)\\
 -i \sin\left(\frac{\pi + 2\epsilon}{4}\right)
\end{pmatrix}; \\ |\psi_3\rangle = 
\begin{pmatrix}
 \cos\left(\frac{\pi - 2\epsilon}{4} + \gamma\right)\\
 -i e^{i\delta}\sin\left(\frac{\pi - 2\epsilon}{4} + \gamma \right)
\end{pmatrix};
\end{equation}

and prepare two identical samples for measurements, $\left(Sample_1, Sample_2\right)$.

Consider a general $\mathcal{PT}$-symmetric Hamiltonian and an associated  $\mathcal{C}$ operator which defines $\mathcal{CPT}$ scalar product:

\begin{equation}
H = H^{\mathcal{PT}} = 
\begin{pmatrix}
r e^{i\beta}; & s; \\
s; & r e^{-i\beta};
\end{pmatrix}; \\  
\mathcal{C} =\frac{1}{\cos\left(\alpha\right)} 
\begin{pmatrix}
i \sin\left( \alpha \right); & 1; \\
1; & -i \sin\left( \alpha \right);
\end{pmatrix};
\end{equation}

where $\sin\left(\alpha\right) = \frac{r}{s}\sin\left(\beta\right)$.
By setting $\sin\left(\alpha\right) = \cos\left( \epsilon \right)$  we make $|\psi_1\rangle$ and $|\psi_2\rangle$ orthogonal in a sense of the $\mathcal{CPT}$ scalar product. Note that $|\psi_1\rangle$ and $|\psi_2\rangle$ \textit{remain} orthogonal by $\mathcal{CPT}$ over time since $\left[\mathcal{C}, \mathcal{PT}\right] = 0$ and $\left[\mathcal{C}, H\right] = 0$ but Hermitian scalar product changes over time since $H^{\dagger} \ne H$.

The $\mathcal{CPT}$ projection operators are: 

\begin{equation}
(|\psi_1\rangle\langle\psi_1|)_{\mathcal{CPT}} = \frac{1}{2\sin\left(\epsilon\right)}\begin{pmatrix}
1 + \sin\left(\epsilon\right); & -i \cos\left(\epsilon\right); \\
-i \cos\epsilon; & -1 + \sin\left(\epsilon\right);
\end{pmatrix}; 
\end{equation}
\begin{equation}
(|\psi_2\rangle\langle\psi_2|)_{\mathcal{CPT}} = \frac{1}{2\sin\left(\epsilon\right)}\begin{pmatrix}
-1 + \sin\left(\epsilon\right); & i \cos\left(\epsilon\right); \\
i \cos\epsilon; & 1 + \sin\left(\epsilon\right);
\end{pmatrix};
\end{equation}

Then, make $|\psi_2\rangle$ and $|\psi_3\rangle$ orthogonal in a  sense of Hermitian scalar product by the Hamiltonian evolution: 
\begin{equation} 
\langle \psi_2| e^{i H^{\dagger} t}e^{-i H t} |\psi_3 \rangle_{Hermitian} = 0
\end{equation}

\begin{equation}
e^{i H^{\dagger} t}e^{-i H t}  = \frac{1}{\cos^2\left(\alpha\right)}\begin{pmatrix}\cos^2\left(\omega t - \alpha\right) + \sin^2\left(\omega t\right); & -2i\sin^2\left(\omega t\right)\sin\left(\alpha\right);\\ 
2i\sin^2\left(\omega \right)\sin\left(\alpha\right); & \cos^2\left(\omega t + \alpha \right) + \sin^2\left(\omega t\right);\end{pmatrix}
\end{equation}

The Hermitian scalar product $(\langle\psi_2|\psi_3\rangle)_{Hermitian} = 0$ becomes zero after the time $\tau$:
\begin{equation}
\tan \left(\omega \tau\right)   = \frac{\sin\left(\epsilon\right)\left(\cos\left(\epsilon \right) \pm \sqrt{2\cos\left(\epsilon\right)\tan\left(\frac{\pi + 2\epsilon}{4}\right)  - 1}\right)}{2\cos\left(\epsilon\right)\tan\left(\frac{\pi + 2\epsilon}{4}\right) - \cos^2\left( \epsilon \right) - 1
 }
\end{equation}

where positive or negative sign depends on the geometry of states and is chosen in such a way that $\tau>0$.

Now, when $(\langle\psi_1|\psi_2\rangle)_{\mathcal{CPT}} = 0$ and $(\langle\psi_2|\psi_3\rangle)_{Hermitian} = 0$, Fig.[\ref{Figure}], make two measurements:
\begin{figure}[h!]
  \includegraphics[width=15cm]{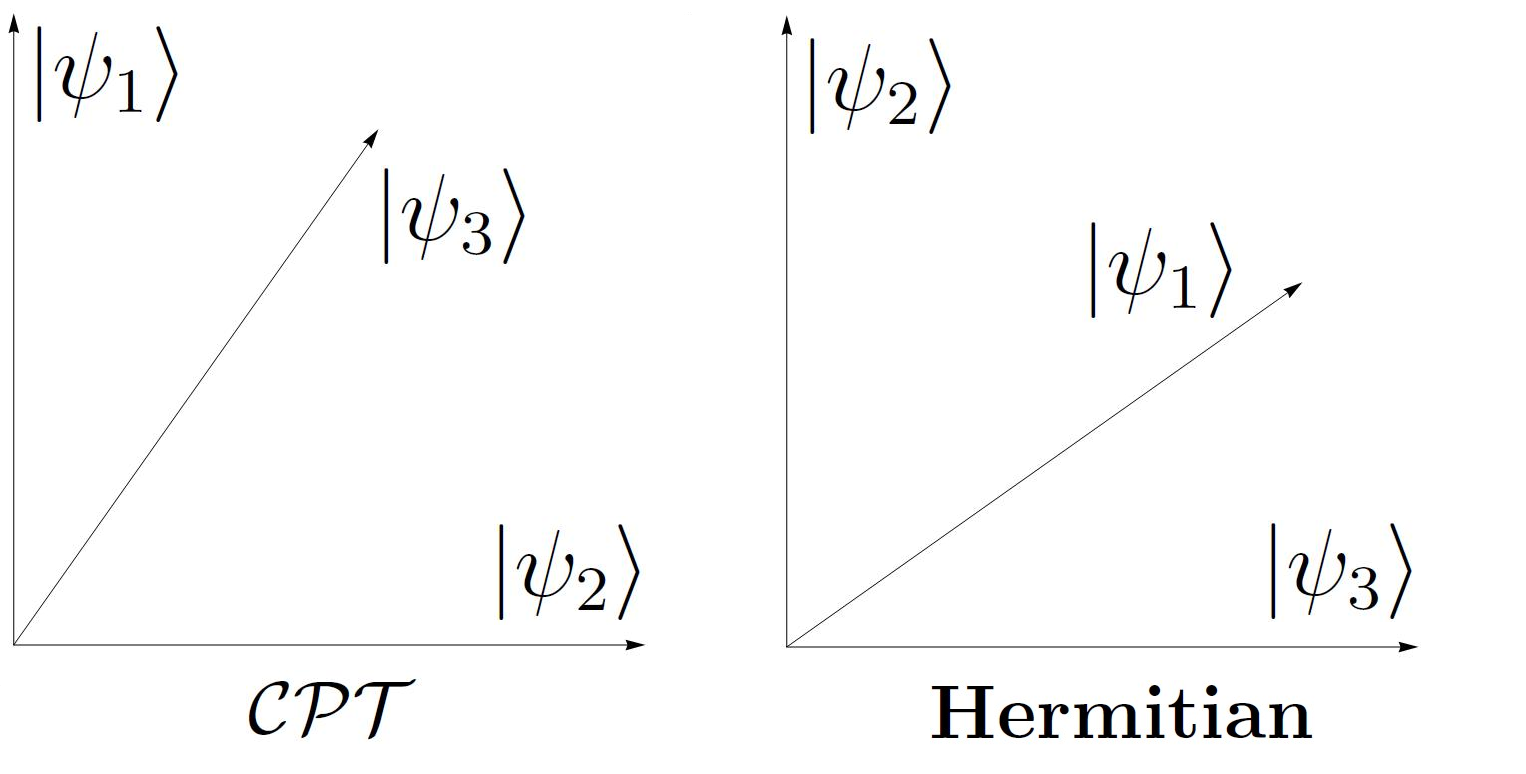}\\
  \caption{$(\langle\psi_1|\psi_2\rangle)_{\mathcal{CPT}}=0$ and $(\langle\psi_2|\psi_3\rangle)_{Hermitian}=0$ after the time $\tau$.}\label{Figure}
\end{figure}
\begin{itemize}
\item Apply the $\mathcal{CPT}$ projection operator on the first sample, 
\begin{equation}
(|\psi_1\rangle\langle\psi_1|)_{\mathcal{CPT}}|Sample_1\rangle = Measurement_1
\end{equation}

\item Apply the Hermitian projection operator on the second sample, 
\begin{equation}
(|\psi_2\rangle\langle\psi_2|)_{Hermitian}|Sample_2\rangle = Measurement_2
\end{equation} 
\end{itemize}

We may get the following results:
\begin{itemize}
\item If $Measurement_1 = 0$, the system is the state $|\psi_2\rangle$ and the problem is solved (since $(\langle\psi_1|\psi_2\rangle)_{\mathcal{CPT}} = 0$)
\item If $Measurement_1 \ne 0$, the system may be in the states $|\psi_1\rangle$ or $|\psi_3\rangle$ and we need to consider the second measurement.
\item If $Measurement_1 \ne 0$ and $Measurement_2 = 0$, the system is in  the state $|\psi_3\rangle$ (since $(\langle\psi_2|\psi_3\rangle)_{Hermitian} = 0$)
\item If $Measurement_1 \ne 0$ and $Measurement_2 \ne 0$, the system is in  the state $|\psi_1\rangle$ (since it is not in the states $|\psi_2\rangle$ and $|\psi_3\rangle$ which we assumed to be pure states)
\end{itemize}

As we shown, \textit{the same Hamiltonian} makes the corresponding $\mathcal{CPT}$ and Hermitian products zero. Another possible solution would be to apply two \textit{different} $\mathcal{PT}$-symmetric Hamiltonians on $Sample_1$ and $Sample_2$ to make $(\langle\psi_1|\psi_2\rangle)_{\mathcal{CPT}}=0$ and $(\langle\psi_2|\psi_3\rangle)_{\mathcal{CPT}}=0$ and then apply two $\mathcal{CPT}$  projections instead of the combined $\mathcal{CPT}$ and Hermitian measurements. 

This scheme can be extended for distinguishing an arbitrary number of pure quantum states. Suppose, we have an arbitrary number of states
\begin{equation}
|\psi_1\rangle, |\psi_2\rangle, \cdots, |\psi_N\rangle
\end{equation}

Prepare $N-1$ samples for measurements and consider three states from the list, $i,j,k \in [1, N]$. Then do the following:
\begin{itemize}
\item Make $(\langle\psi_i|\psi_j\rangle)_{\mathcal{CPT}}=0$ and $(\langle\psi_j|\psi_k\rangle)_{Hermitian}=0$ (or $(\langle\psi_i|\psi_j\rangle)_{\mathcal{CPT}}=0$ and $(\langle\psi_j|\psi_k\rangle)_{\mathcal{CPT}}=0$)
\item Apply the $\mathcal{CPT}$ projection operator on the first sample, 
\begin{equation}
(|\psi_i\rangle\langle\psi_i|)_{\mathcal{CPT}}|Sample_1\rangle = Measurement_1 
\end{equation}
\item Apply the Hermitian (or $\mathcal{CPT}$) projection operator on the second sample, 
\begin{equation}
(|\psi_j\rangle\langle\psi_j|)_{Hermitian/\mathcal{CPT}}|Sample_2\rangle = Measurement_2
\end{equation}
\end{itemize}  
We find a state of the system if one of the measurements is zero, or exclude two possibilities from our list. By applying this procedure several times on the remaining states from our list, we exhaust all possibilities and find the state of the system assuming it's in the pure state.

The question on whether this scheme could be extended to mixed states we refer to our future research.

\section{Acknowledgments}
Y.B. was supported by the UCCS Graduate Research Award.

\end{document}